\documentclass[12pt,a4paper]{scrartcl}
\usepackage[utf8]{inputenc}
\usepackage[T1]{fontenc}
\usepackage{lmodern}
\usepackage[british]{babel}
\usepackage{amsmath}
\usepackage{amssymb}
\usepackage{amsfonts,ulem,bbm}
\usepackage{color}
\usepackage{float}
\usepackage{appendix}
\usepackage[pdftex]{graphicx}
\usepackage{epstopdf}
\usepackage{grffile}
\usepackage{hyperref}
\usepackage{textcomp}
\usepackage{subfigure}

\newcommand{\tr}{\ensuremath{\mathrm{tr}}}
\newcommand{\pf}{\ensuremath{\mathrm{Pf}}}
\newcommand{\re}{\ensuremath{\mathrm{Re}}}
\newcommand{\dt}{\ensuremath{\mathrm{det}}}
\newcommand{\sign}{\ensuremath{\mathrm{sign}}}
\newcommand{\Rnum}[1]{\uppercase\expandafter{\romannumeral #1\relax}}

\newcommand{\Nt}{\ensuremath{N_\tau}}

\newcommand{\bcd}{\beta_c^\textrm{dec}}
\newcommand{\SU}{\ensuremath{\mathrm{SU}}}
\newcommand{\Z}{\ensuremath{\mathbb{Z}}}

\newcommand{\RS}{\ensuremath{\mathbb{R}^3\times\mathbb{S}^1}}
\newcommand{\adjQCDc}{\ensuremath{\mathrm{adjQCD}_{\mathbb{R}^3\times \mathbb{S}^1}}}
\newcommand{\Eqref}[1]{Eq.~(\ref{#1})}

\newcommand{\arxv}[2]{\href{http://arxiv.org/abs/#2}{\texttt{{[}#1\,]}}}

\title{
  {\vspace{-20mm}\normalsize
   \hfill\parbox[b][30mm][t]{35mm}{\textmd{MS-TP-14-32}}}\\[-18mm]
Compactified $\mathcal{N}=1$ supersymmetric Yang-Mills theory on the lattice:\\
 Continuity and the disappearance of the deconfinement transition}
\author{G.~Bergner\\
\textit{\large Universit\"at Bern, Institut für Theoretische Physik}\\
\textit{\large Sidlerstr.~5, CH-3012 Bern, Switzerland}\\
\textit{\large E-mail: bergner@itp.unibe.ch}\\[5mm]
 S.~Piemonte\\
\textit{\large Universit\"at M\"unster, Institut f\"ur Theoretische Physik}\\
\textit{\large Wilhelm-Klemm-Str.~9, D-48149 M\"unster, Germany}\\
\textit{\large E-mail:  spiemonte@uni-muenster.de}
}

\date{}

\begin{document}

\maketitle

\begin{abstract}
Fermion boundary conditions play a relevant role in revealing the confinement mechanism of $\mathcal{N} = 1$ 
supersymmetric Yang-Mills theory with one compactified space-time dimension. A deconfinement phase transition occurs for a sufficiently small compactification radius, equivalent to a high temperature in the thermal theory where antiperiodic fermion boundary conditions are applied. Periodic fermion boundary conditions, on the other hand, are related to the Witten index and confinement is expected to persist independently of the length of the compactified dimension. We study this aspect with lattice Monte Carlo simulations for different values of the fermion mass parameter that breaks supersymmetry softly. We find a deconfined region that shrinks when the fermion mass is lowered. Deconfinement takes place between two confined regions at large and small compactification radii, that would correspond to low and high temperatures in the thermal theory. At the smallest fermion masses we find no indication of a deconfinement transition. These results are a first signal for the predicted continuity in the compactification of supersymmetric Yang-Mills theory.
\end{abstract}
\section{Introduction}
Gauge theories with adjoint fermions (adjQCD) have interesting thermodynamical properties and the study of their phase transitions provides a deeper understanding of strong interactions at finite temperature. The $\mathcal{N} = 1$
supersymmetric Yang-Mills theory (SYM) is a special case among adjQCD theories with a different number of fermions. One
main motivation to study this theory has been its role as gauge part of extensions of the standard model. The phase diagram of $\mathcal{N} = 1$ SYM has been analysed at finite temperatures in a previous publication \cite{Bergner:2014saa}. Supersymmetry is broken at non-zero temperature as a consequence of the different thermal statistics of fermions and bosons. In this contribution we focus our attention on the phase transitions of the compactified SYM with periodic fermion boundary conditions. Supersymmetry is preserved in this theory and is expected to have a considerable influence on the phase diagram.

Confinement and fermion condensation are the two relevant phenomena of QCD-like theories regardless of whether the fermions are in the fundamental or adjoint representation. At low temperatures the theory is in a confined phase with
colourless strongly bound particles and unbroken centre symmetry. Chiral symmetry is broken by a non-vanishing fermion
condensate. At high temperatures there is a phase transition to a deconfined phase with spontaneously broken centre
symmetry. The chiral condensate melts away leading to a restoration of chiral symmetry. However, the deconfinement
transition is only a mild crossover in QCD and other theories with fermions in the fundamental representation, due to the
explicit breaking of centre symmetry by the quark action. By contrast, the transition from the confined to the deconfined
phase is a true phase transition in adjQCD models for any value of the fermion mass and in the massless limit chiral
symmetry restoration defines a second one that can have a different critical temperature.

The picture changes completely when the boundary conditions of fermions are changed from thermal,
i.e.\ antiperiodic, to periodic. The path integral of the compactified theory on \RS\ with periodic fermion boundary conditions (\adjQCDc) corresponds to a twisted partition function instead of the usual thermal partition function $Z=\tr[ e^{-H/T}]$. For SYM this twisted partition function represents the Witten index \cite{Witten:1982df}
\begin{equation}
\tr[ (-1)^F e^{-H/T}]= \sum_{\substack{\textrm{boson}\\\textrm{states}}} e^{(-E_n/T)} -\sum_{\substack{\textrm{fermion}\\\textrm{states}}} e^{(-E_n/T)} = \int_\textrm{PBC} \mathcal D\psi \mathcal D A_\mu e^{-S[\psi,A]}\; ,
\end{equation}
where the fermion number $F$ is odd for a fermionic state and otherwise even. 
If the same periodic boundary conditions (PBC) are applied in a compactified theory for adjoint quark and gauge fields,
then an interesting interplay exists between bosonic and fermionic degrees of freedom which avoids, in case of SYM, an
explicit supersymmetry breaking in contrast to the thermal case. The fermionic contributions can cancel the confining
potential of the gauge bosons leading to a restoration of centre symmetry. In SYM there is a cancellation
to all orders in the perturbative expansion and a centre stabilisation by non-perturbative semi-classical contributions \cite{Unsal:2008ch,Poppitz:2012sw,Unsal:2010qh}. A complicated breaking
pattern is obtained for general $\SU(N_c)$ gauge groups, where additional phases appear when only parts of the $\Z_{N_c}$
centre symmetry are broken \cite{Myers:2009df}. Such phases were also found in Yang-Mills theory extended by adjoint
Polyakov loop terms, which are similar to the heavy quark limit of \adjQCDc\ \cite{Myers:2007vc}.

There are different theoretical concepts related to \adjQCDc . The first of them is the Hosotani mechanism
\cite{Hosotani:1983xw}, the possibility that a partial breaking of the gauge symmetry in the compactified theory allows
to interpret the gauge field of the compactified direction as a Higgs field in a lower dimensional theory. This 
gauge-Higgs unification plays an important role in extensions of the standard model with an extra dimension.

A further motivation for the investigations of \adjQCDc\ is the large $N_c$ volume independence of gauge theories, known
as Eguchi-Kawai reduction \cite{Eguchi:1982nm}. This reduction implies an equivalence between the full four-dimensional
gauge theory and a simple single site matrix model in the large $N_c$ limit. However, volume independence is known to
fail for pure Yang-Mills due to the spontaneous breaking of centre symmetry driven by the compactification
\cite{Bhanot:1982sh,Kovtun:2007py}. Adding adjoint fermions to the model (adjoint Eguchi-Kawai models) can in principle
resolve the centre symmetry breaking keeping the large $N_c$ volume independence intact
\cite{Kovtun:2007py,Hollowood:2009sy}.

The dependence of the ground state on the parameters of the theory can be determined from the effective potential.
A perturbative loop expansions of the effective potential is characterised by powers of the coupling constant $g^2$ and
a complete semi-classical expansion adds non-perturbative contributions, that typically come with 
exponentials of the coupling like $e^{-1/g^2}$.
The one-loop approximation of the effective potential for pure Yang-Mills theory (YM) predicts the deconfined phase 
with spontaneously broken centre symmetry at high temperatures, and in QCD, with fermions in the fundamental 
representation, the explicit breaking of centre symmetry is reproduced at one-loop order. The applicability of semi-classical 
methods in QCD at lower temperatures and towards the deconfinement transition is limited.
With intact supersymmetry there is an exact cancellation between fermionic and bosonic perturbative contributions in the loop 
expansion of the effective potential. The non-perturbative semi-classical effects are the dominant part of the effective
potential \cite{Poppitz:2012sw}. Compactified SYM is thus an interesting theory for the investigation of semi-classical non-perturbative contributions.

In this work we consider compactified $\SU(2)$ SYM on \RS\ with periodic (PSYM) and thermal\footnote{Thermal means antiperiodic boundary conditions for fermion fields, but periodic ones for gauge fields.} (TSYM) boundary conditions and investigate different aspects of the deconfinement transition.
For the first time we perform lattice simulations of this theory that capture in principle all perturbative 
and non-perturbative contributions. In particular we are interested in the differences  with respect to the thermal deconfinement transition
that we have studied in our previous investigations.

\adjQCDc was the subject of earlier investigations on the lattice in the context of the Hosotani mechanism \cite{Cossu:2009sq, Cossu:2013ora}. 
Adjoint Eguchi-Kawai models reduced to a single lattice site or small volume were investigated in \cite{Hietanen:2009ex,Hietanen:2010fx,Bringoltz:2011by,Catterall:2010gx,Cunningham:2013wha,Azeyanagi:2010ne}.
Recently a method for numerical simulations based on the semi-classical 
analysis was tested in \cite{Anber:2013doa,Tatsuhiro:2014sw}. 
\section{Compactified supersymmetric Yang-Mills theory}
\begin{figure}
\centering
\includegraphics[width=0.69\textwidth]{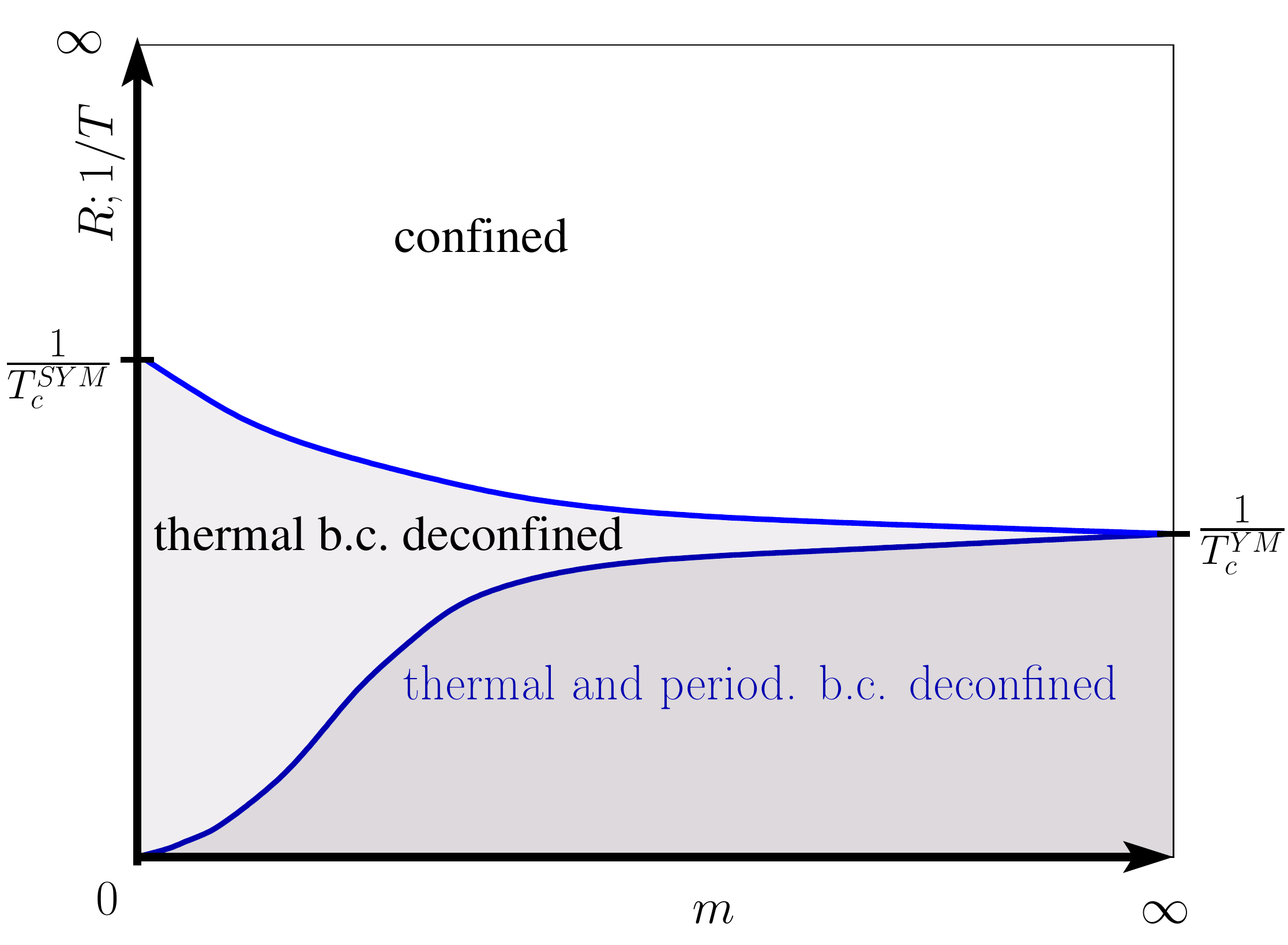}
\caption{The phase diagram of SYM according to the theoretical predictions \cite{Unsal:2010qh}.
In the theory with thermal, i.~e.\ antiperiodic, fermion boundary conditions the critical 
deconfinement radius $R$ is the inverse of the critical temperature $T$.
The thermal theory has a larger critical deconfinement radius
than the one with periodic fermion boundary. The dark shaded part indicates
the deconfined region for both theories.
}
\label{fig:theoryphase1}
\end{figure}
\begin{figure}
\centering
\includegraphics[width=0.69\textwidth]{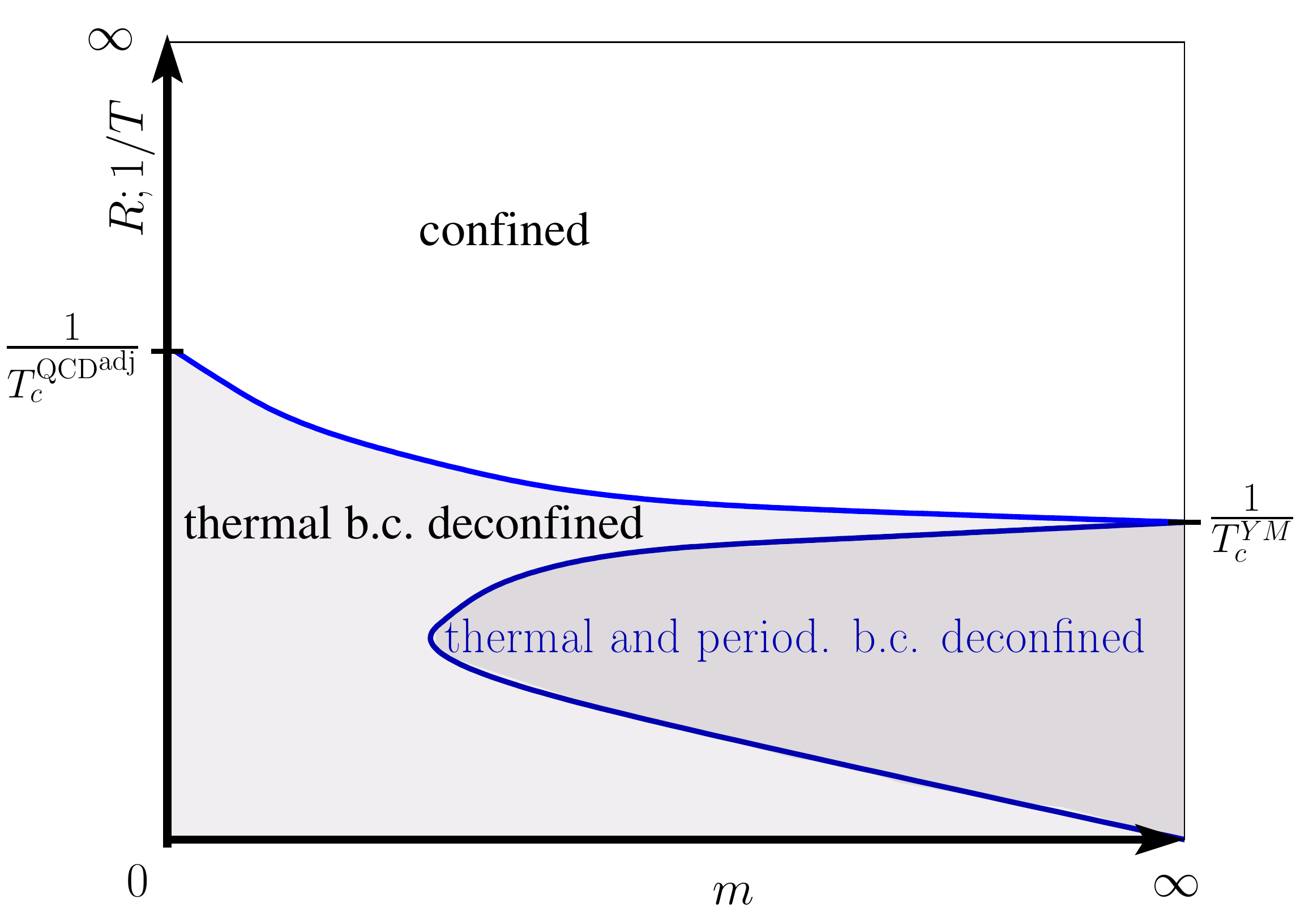}
\caption{The same phase diagram as in Fig.~\ref{fig:theoryphase1} now for a larger number of Majorana fermions ($N_f > 1$), but still outside the conformal window.
}
\label{fig:theoryphase2}
\end{figure}
\begin{figure}[t]
\centering
\includegraphics[width=0.69\textwidth]{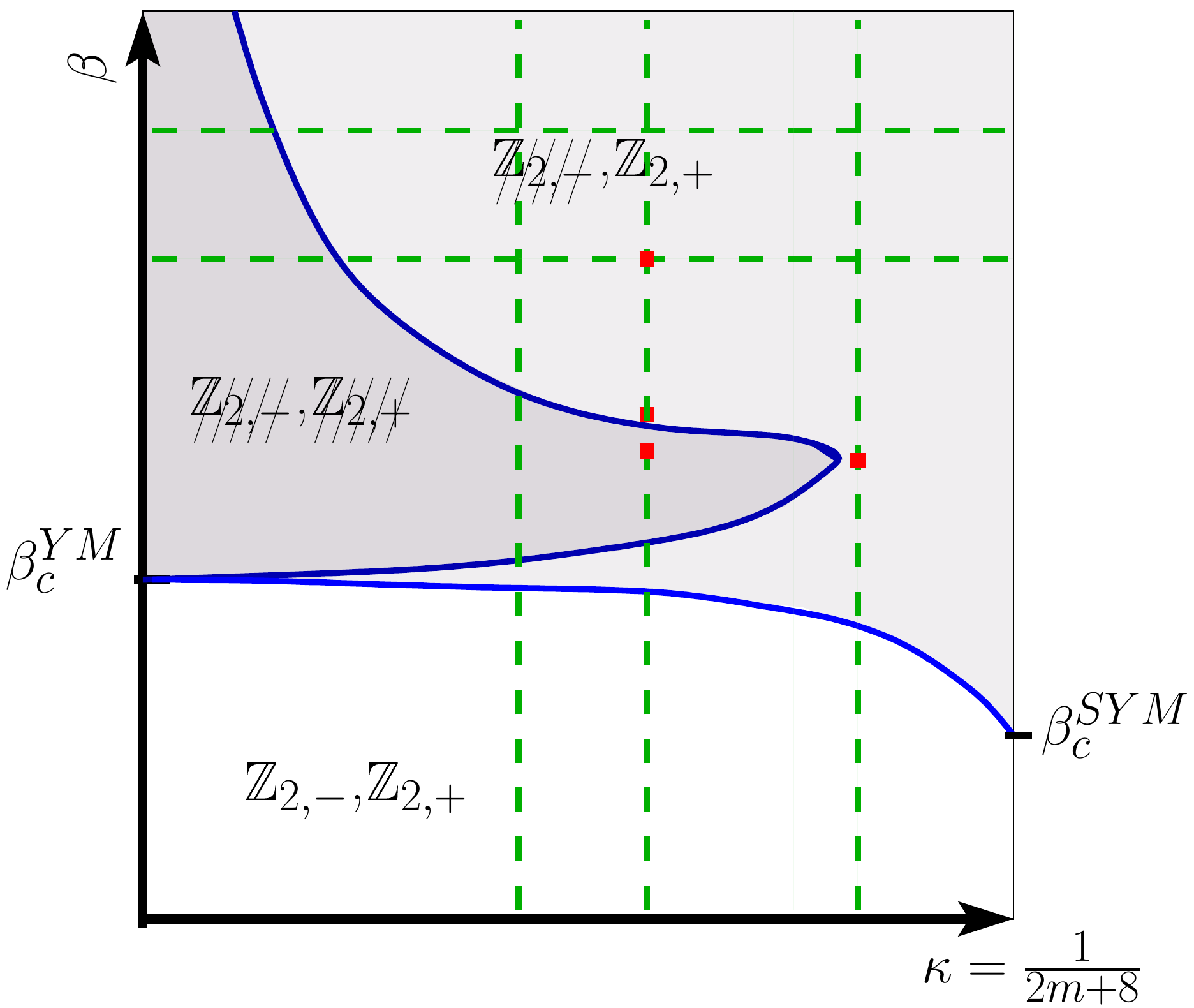}
\caption{The phase diagram found in lattice simulations of SYM: 
$\mathbbm{Z_{2,+}}$/$\xout{\mathbbm{Z_{2,+}}}$ ($\mathbbm{Z_{2,-}}$/$\xout{\mathbbm{Z_{2,-}}}$) stands for confinement/deconfinement in the theory with periodic (antiperiodic) fermion boundary conditions.
The green lines show the scans of the parameter range performed in the simulations on an $N_\tau=4$ lattice. 
The red dots are the position of precise checks of the phases with the histogram of the Polyakov line at different volumes. A large value of $\beta$ corresponds to a small compactification radius $R$.
}
\label{fig:summaryresults}
\end{figure}
In a previous publication \cite{Bergner:2014saa} we have analysed the thermal phase transitions of \SU(2) SYM theory.
We start with a brief review of these results.

The Euclidean on-shell action in the continuum is
\begin{equation}
\label{continuum_action}
S(g,m) =  \int d^4 x \left\{\frac{1}{4} (F_{\mu\nu}^a F_{\mu\nu}^a) 
+ \frac{1}{2}\sum_{n_f=1}^{N_f}\bar{\lambda}_{n_f} (\gamma^\mu D_\mu + m)\lambda_{n_f}\right\}\, ,
\end{equation}
where $F_{\mu\nu}$ is the field strength tensor and $D_\mu$ the gauge covariant derivative
\begin{equation}
 F_{\mu\nu}=\partial_\mu A_\nu- \partial_\nu A_\mu+i g[A_\mu, A_\nu];\quad (D_\mu\lambda)^a=\partial_\mu\lambda^a-gf_{abc} A^b_\mu\lambda^c\; ,
\end{equation}
with the structure constants $f_{abc}$ of the gauge group.

The fields $\lambda$ represent Majorana fermions in the adjoint representation of the gauge group.
There is only one Majorana fermion in SYM ($N_f=1$) and it is the supersymmetric partner of the gluon called gluino. The additional non-zero gluino mass term leads to a soft supersymmetry breaking. Full supersymmetry is recovered in
the limit where the renormalised gluino mass vanishes.

The theory is confined at low temperature, confirmed by the linear rise of the static quark-antiquark potential in lattice simulations. The bound state spectrum has been investigated in earlier studies of our collaboration \cite{Bergner:2012rv,Bergner:2013nwa}.

Chiral $\mathrm{U}(1)_\mathrm{R}$ symmetry has a non-trivial breaking pattern in this theory.
This symmetry is broken by an anomaly as in one flavor QCD, but a discrete $\Z_{2N_c}$ subgroup is left intact in theories with fermions in the adjoint representation. This remaining symmetry is spontaneously broken down to $\Z_2$ by a non-vanishing expectation value of the gluino condensate.

A deconfined phase with restored $\Z_{2N_c}$ chiral symmetry is expected at sufficient high temperatures. There are no simple theoretical connections between centre and chiral symmetry, therefore two phase transitions can occur independently at two different temperatures in SYM and other theories with adjoint fermions.
Chiral and deconfinement phase transitions have been found to occur roughly at the same temperature in our previous lattice simulations of SU(2) SYM within our current precision, leaving the question on whether there exists a dynamical hidden link between them.

The present work is focused on understanding how the deconfinement phase transition is affected by the fermion boundary conditions. In thermal SYM we have found that, if $\Nt$ is fixed, the deconfinement transition appears at a lower value of bare coupling $\beta$ when the gluino mass is decreased. With an appropriate scale setting, the temperature of the deconfinement transition in TSYM is lower than in pure YM.

These observations of the thermal transition are opposed to what we expect to find in our new simulations of PSYM (see Fig.\ \ref{fig:theoryphase1} and \ref{fig:theoryphase2}): when the gluino mass decreases the critical compactification radius $R$, which can be identified with an inverse temperature, decreases.
The critical $R$ is even expected to vanish in the supersymmetric limit ($m\rightarrow 0$) and the deconfinement transition should completely disappear in PSYM without the soft supersymmetry breaking of the mass term.
This supersymmetric limit is approached smoothly by the predicted transition line \cite{Unsal:2010qh}.
This implies that at very small $R$ the theory is confined only for very small values of the gluino mass $m$.

The reduced deconfined region in the phase diagram is induced by the adjoint fermions with periodic boundary conditions. 
Therefore a larger $N_f$ is expected to increase the confined region even further,
according to \cite{Unsal:2010qh} the deconfinement transition completely disappears already at a finite $m$
for $N_f>1$.\footnote{It is assumed that the theory is still outside the conformal window. Note, however, that already $N_f=2$ could be conformal \cite{Athenodorou:2013eaa}.}
In this case the confined region at large $R$ is connected to a confined region at small $R$. At very small $R$ the theory is
confined up to a large value of $m$ that tends to infinity as $R$ goes to zero. At an infinite mass there is, of course,
always the deconfined region of pure YM.

The results of our lattice SYM simulations are summarised in Fig.~\ref{fig:summaryresults} and represented in terms of the parameters $\beta=\frac{2 N_c}{g^2}$ and $\kappa=\frac{1}{2(m+4)}$. The scale of our simulations depends on the gauge coupling $g$, in particular the lattice spacing is an exponentially decreasing function of $\beta$. At a fixed temporal extend $\Nt$ of the lattice the larger critical parameter $\bcd$ is hence
equivalent to a smaller critical compactification radius or a larger critical temperature.

The absence of the deconfinement transition is confirmed in the supersymmetric limit.
At finite $m$ the results are, however, in favour of the picture expected for larger values of $N_f$. 
Within the limited volume and mass range accessible in our simulations we find no evidence for a deconfinement 
transition below a certain value of the bare mass parameter and a shrinking deconfined region at smaller values of 
$R$. Implications and limitations of these findings are discussed in the conclusions, Sec.~\ref{sec:conclusions}.

After a short introduction of our methods, already applied in \cite{Bergner:2014saa}, we summarise our numerical results providing evidence for this scenario in the following sections.
We have done scans in the bare parameter space for many different $\beta$ and fixed bare mass parameters $\kappa$.
An important point in this analysis is the investigation of finite size effects.
The theory at small $R$ has an almost flat effective potential for the Polyakov loop, leading to large fluctuations and 
autocorrelations of this observable. 
The effect appears similar to the broadening induced by the tunnelling between the two $Z_2$ minima in the deconfined phase at
smaller volumes. Only in a comparison of different volumes it is possible to 
discriminate the broad distribution in the confined region at small $R$ from the broadening of the distribution by tunnelling due to finite volume effects in the deconfined region. 
A comparison of the Polyakov loop histograms for different volumes provides an estimate of the finite volume effects. 
We have performed this study at certain points in the phase diagram as sketched in Fig.~\ref{fig:summaryresults}.
\section{Lattice simulations}
In our simulations we have used a tree-level Symanzik improved gauge action
and Wilson-Dirac fermions
\begin{align}
 S = \sum_x\! \re\, \tr \left[\frac{\beta}{N_c}\sum_{\mu \neq \nu}
\left(\frac{5}{3}  P_{\mu\nu}(x) - \frac{1}{12} R_{\mu\nu}(x)\right)\right]
+\frac{1}{2}\sum_{n_f;x,y} \bar{\lambda}_{n_f}(y) D_W[V_\mu](y,x) \lambda_{n_f}(x),
\end{align}
where $P_{\mu\nu}(x)$ is the plaquette and $R_{\mu\nu}(x)$ the rectangle 
of gauge links introduced as an improvement of the standard Wilson gauge action.
$U_\mu(x)$ and $V_\mu(x)$ denote the link variables in
the fundamental and in the adjoint representation, respectively. The adjoint
links $V_\mu(x)$ are related to the fundamental links $U_\mu(x)$ through the
well-known formula
\begin{equation}
 V_\mu(x)_{ab} = 2\, \tr{(U_\mu(x)^\dag \tau_a^F U_\mu(x) \tau_b^F)}\, ,
\end{equation}
where the generators in the fundamental
representation $\tau_a^F$ are normalised such 
that $ \tr (\tau_a^F \tau_b^F) = \frac{1}{2} \delta_{ab}$.
The action of the Wilson-Dirac operator $D_W$ on the gluino field $\lambda$
is defined as
\begin{equation}
 D_W(x,y)\lambda(y) =\! \lambda(x) - \kappa \sum_{\mu}\! 
\left[ (1-\gamma_\mu) V_\mu(x) \lambda(x+\mu) 
+ (1+\gamma_\mu) V_\mu(x-\mu)^\dag \lambda(x-\mu)\right].
\end{equation}

On the lattice chiral symmetry and supersymmetry are explicitly broken.
The tuning of the bare gluino mass $m$ 
is enough in supersymmetric Yang-Mills theories to recover both symmetries in the continuum limit \cite{Curci:1986sm}.
The chiral limit can be reached approaching the point where the adjoint pion mass, defined in a
partially quenched setup \cite{Munster:2014cja}, vanishes. 

There is a sign problem in the lattice discretised theory if the total number of Majorana fermions $N_f$ is odd, as in the case of SYM. Fermions are integrated out to perform numerical simulations and the 
result is the Pfaffian of the Wilson-Dirac operator
\begin{equation}
 Z = \int D U\ \pf ( C D_W )^{N_f} \exp{(-S_g)}.
\end{equation}
The modulus of the Pfaffian is the square root of the
determinant, leaving an additional sign factor for odd $N_f$
\begin{equation}
 \pf ( C D_W ) = \sign(\pf ( C D_W )) \sqrt{\dt(D_W)}.
\end{equation}
The sign of the Pfaffian is positive in the continuum limit, but on the 
lattice configurations with negative sign can occur and the probability that the sign changes during a Monte Carlo simulation increases at smaller gluino masses for fixed lattice spacing. 
In the compactified theory with periodic boundary conditions sign changes are more likely compared to the theory with thermal boundary conditions. In our current investigations we avoid entering the region with a relevant number of sign changes 
by keeping the gluino mass far enough from its critical value. 
This is checked by a measurement of  the Pfaffian signs on a subset of configurations for the runs
with the most critical parameters using the method introduced in
\cite{Bergner:2011zp}.
Note that with periodic boundary conditions the problem becomes already 
relevant at $\kappa\approx 0.19$.

As in our previous investigations \cite{Bergner:2014saa} the simulations are done with the
RHMC algorithm. Towards the supersymmetric limit (vanishing renormalised gluino mass), 
the cost of the RHMC trajectory increases drastically. This problem 
is common to all simulations with dynamical fermions and becomes even more significant with periodic boundary conditions.
Therefore the limit of small gluino masses can only be reached at a high cost.
\section{Numerical results for compactified supersymmetric Yang-Mills theory}
In this section we provide numerical evidence for the following facts for the compactified \SU(2)  SYM theory on \RS :
as expected, there is no difference between PSYM and TSYM at small $\beta$ where both are in
the low temperature $T$ (or large radius $R$) confined phase.
Moving towards the deconfinement transition line, we observe that the difference in the fermion boundary conditions becomes significant even at a rather large gluino mass. At large $\beta$ and a wide range of the bare mass 
parameter $\kappa$ we find a phase with unbroken centre symmetry,
similar to the ``re-confined phase'' in \cite{Cossu:2013ora}. 
At larger gluino masses there is a clear signal for spontaneously broken centre symmetry and a deconfined phase between 
the re-confined phase and the confined phase at small $\beta$.
The two confined phases are connected: at lower gluino masses the signal for deconfinement vanishes.
The deconfined phase close to the pure Yang-Mills limit
shrinks towards larger values of $\beta$ leading to a sharp transition when $\kappa$ is increased.

The volume averaged Polyakov loop, 
\begin{equation}
P_L = \frac{1}{V} \sum_{\vec{x}} 
\textrm{Tr}\left\{ \prod_{t=0}^{\Nt} U_4(\vec{x},t)\right\}\;,
\label{eq:PL}
\end{equation}
is an order parameter of the deconfinement transition.
The constraint effective potential of the Polyakov loop has either a minimum at zero, in the confined phase, or two degenerate
minima representing the spontaneously broken $\Z_2$ centre symmetry in the deconfined phase.
The histogram of the Polyakov loop is another representation of the constraint effective Potential.
The distribution is either centred around a maximum at $P_L=0$ in the confined phase, or around two symmetric non-zero peaks in the 
deconfined phase. There is a finite tunnelling rate in the deconfined phase between the two minima corresponding to these peaks, that is suppressed in the infinite volume limit. 
The broad distribution of the Polyakov loop induced by tunnelling is hard to distinguish from a signal for confinement.
To identify the different phases it is necessary to compare the histograms of simulations at different volumes, in particular for the confined phase at large $\beta$ that is characterised by a rather broad distribution of the Polyakov loop. 
Due to this broad distribution it is expected that the signal for the transition point can only be identified at rather large volumes.
In this work we study the expectation value of the modulus of the volume averaged Polyakov loop, $\langle|P_L|\rangle$, since it provides a clearer signal for the deconfinement at finite volumes.
\begin{figure}[t]
\centering
\includegraphics[width=0.7\textwidth]{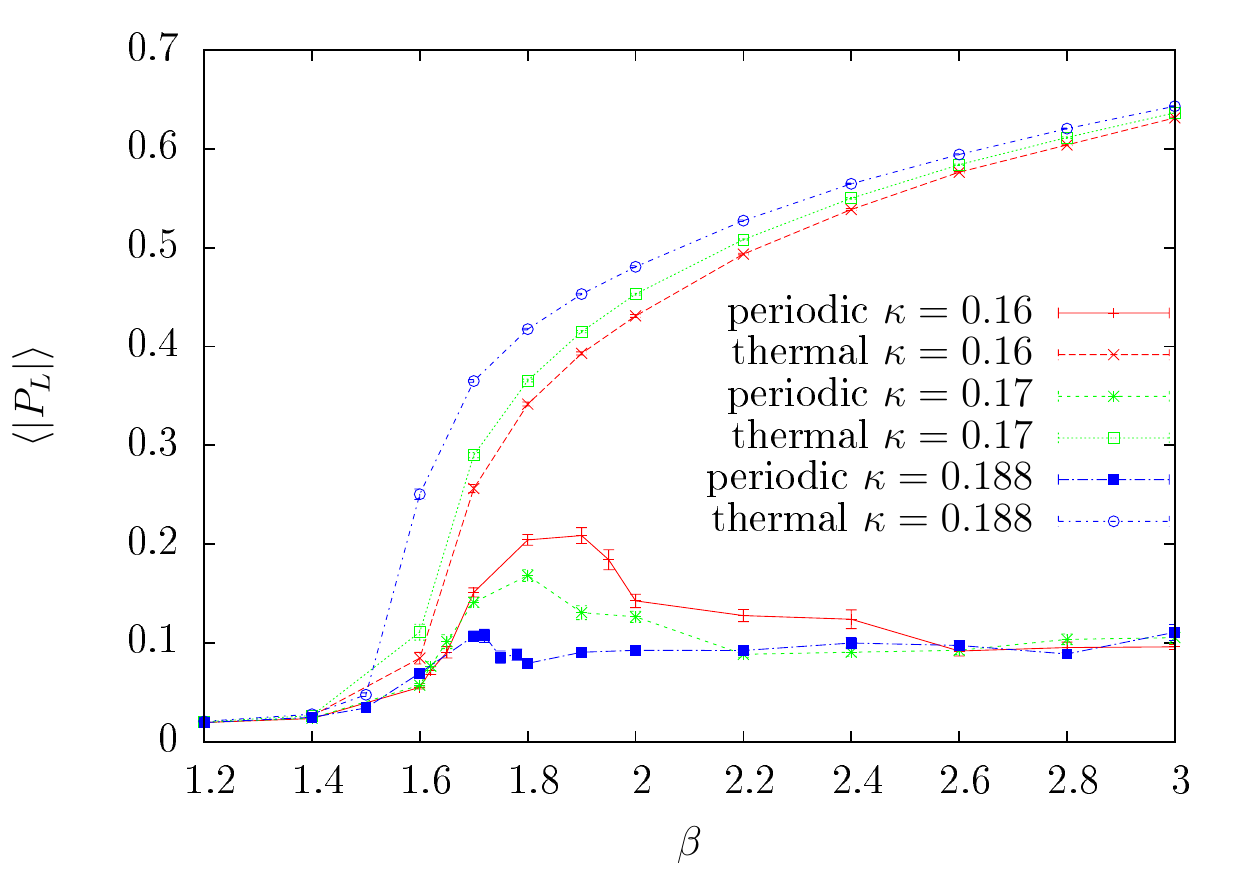}
\caption{
The measured modulus of the volume averaged Polyakov loop $|P_L|$ (\Eqref{eq:PL}) in scans of the inverse bare coupling $\beta$ at 
a fixed value of the bare mass parameter $\kappa$ on an $\Nt\times N_s^3=4\times 8^3$ lattice.
With thermal (antiperiodic) fermion boundary conditions the signal for the deconfinement transition, that moves towards lower $\beta$ at larger $\kappa$, is clearly visible in this picture. In the theory 
with periodic boundary fermion conditions, on the other hand, the picture is completely different. A larger $\langle|P_L|\rangle$ are obtained only at intermediate values of $\beta$.}
\label{fig:bscan}
\end{figure}
\begin{figure}[t]
\centering
\subfigure[]{\includegraphics[width=0.49\textwidth]{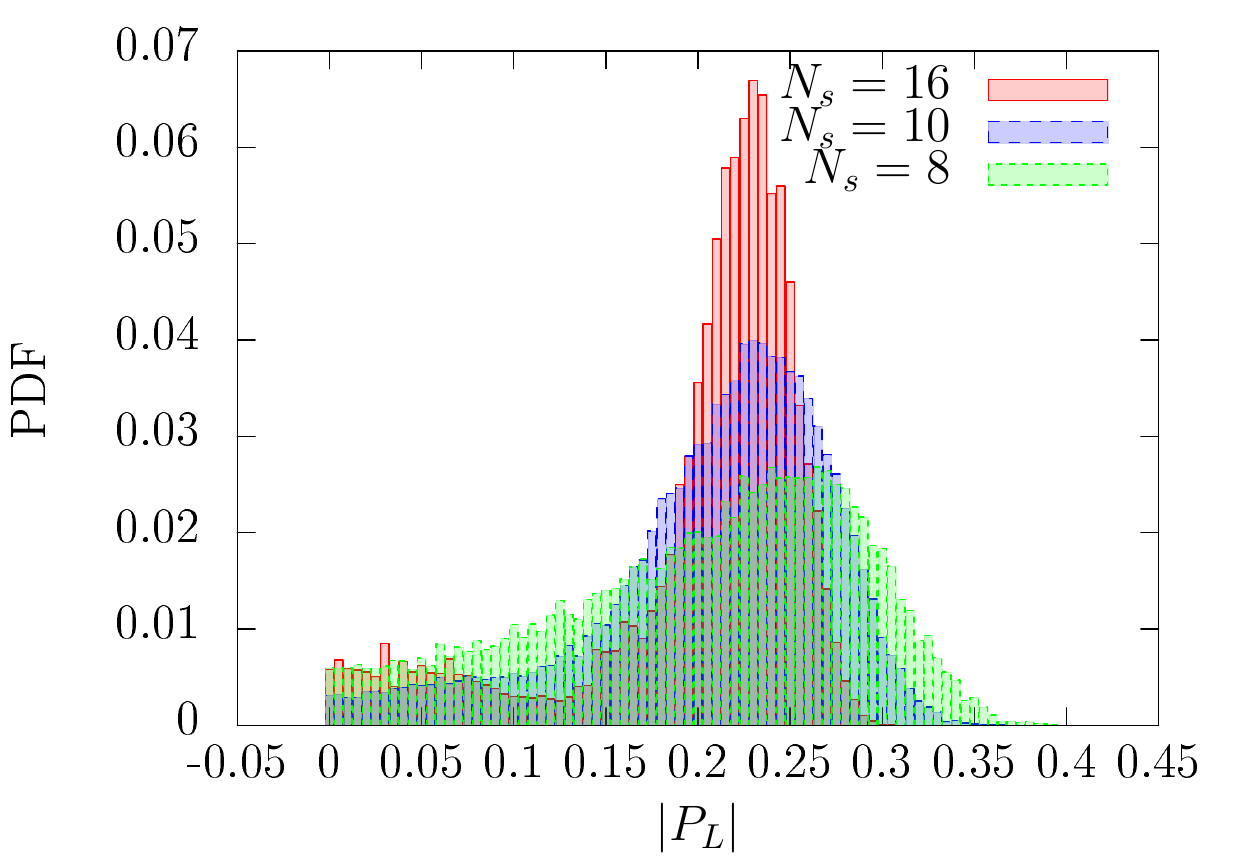}}
\subfigure[]{\includegraphics[width=0.49\textwidth]{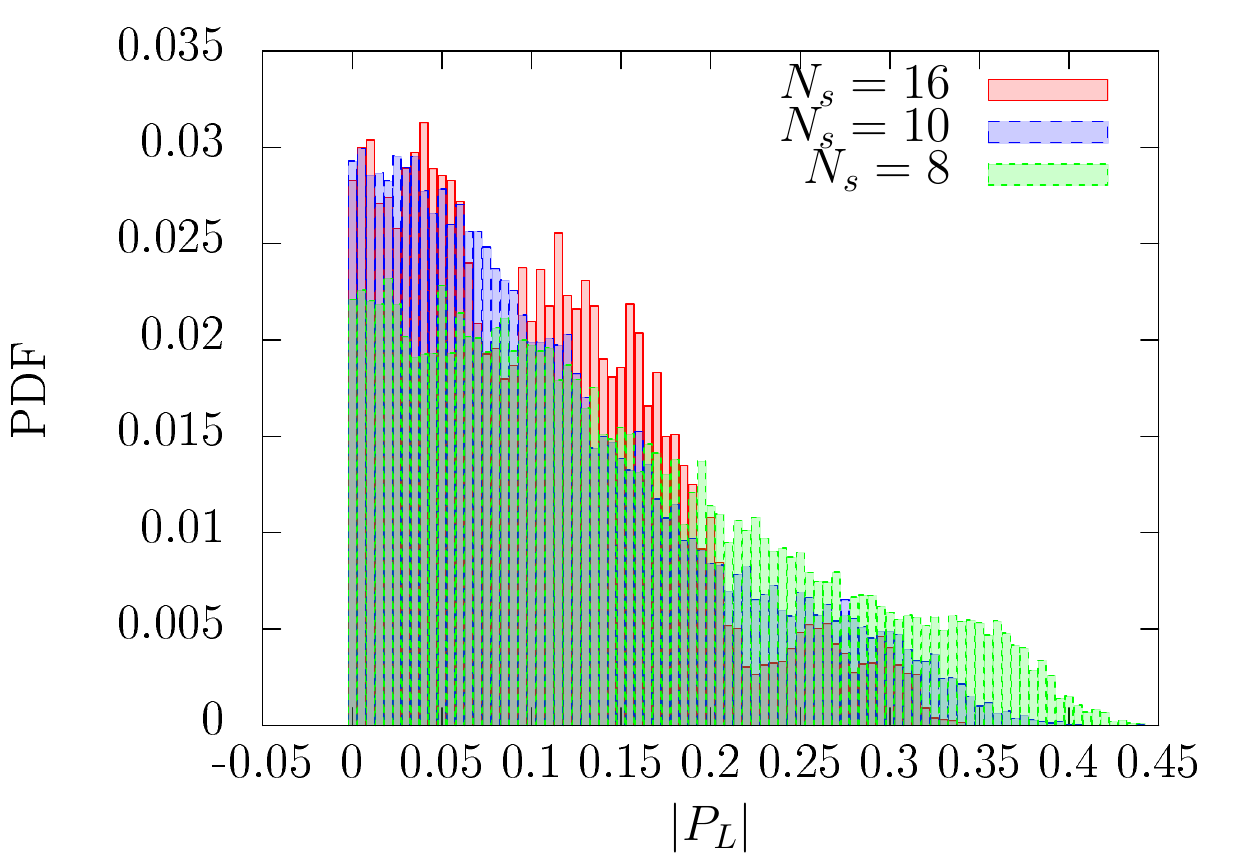}}
\subfigure[]{\includegraphics[width=0.49\textwidth]{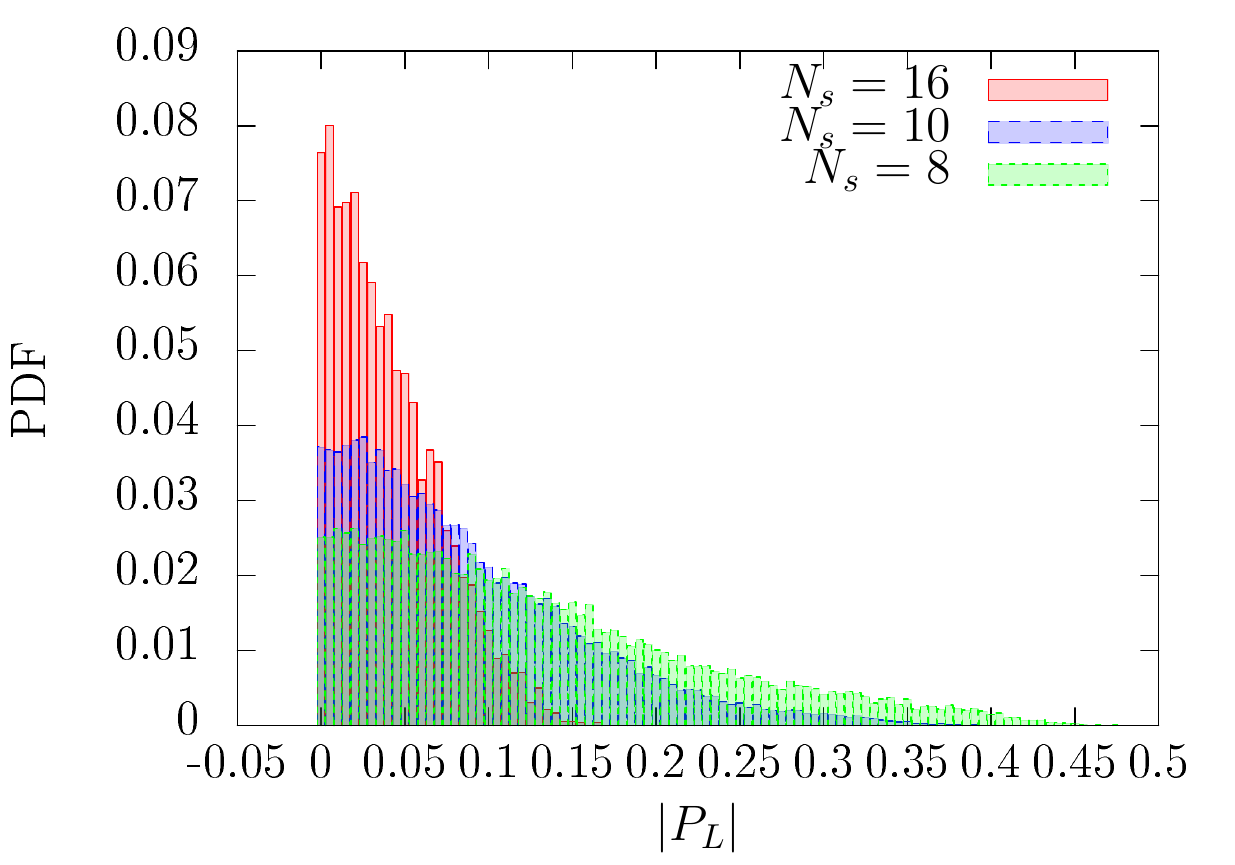}}
\caption{The histograms of $|P_L|$ for $\kappa=0.16$ from simulations on
 $4\times N_s^3$ lattices. The different volumes $N_s^3$ are compared to show the finite size effects.
The theory changes from the deconfined phase at $\beta=1.8$ (a) to a confined phase at
$\beta=2.0$ (b) and $\beta=2.2$ (c).
}
\label{fig:hist}
\end{figure}
\begin{figure}[t]
\centering
\includegraphics[width=0.7\textwidth]{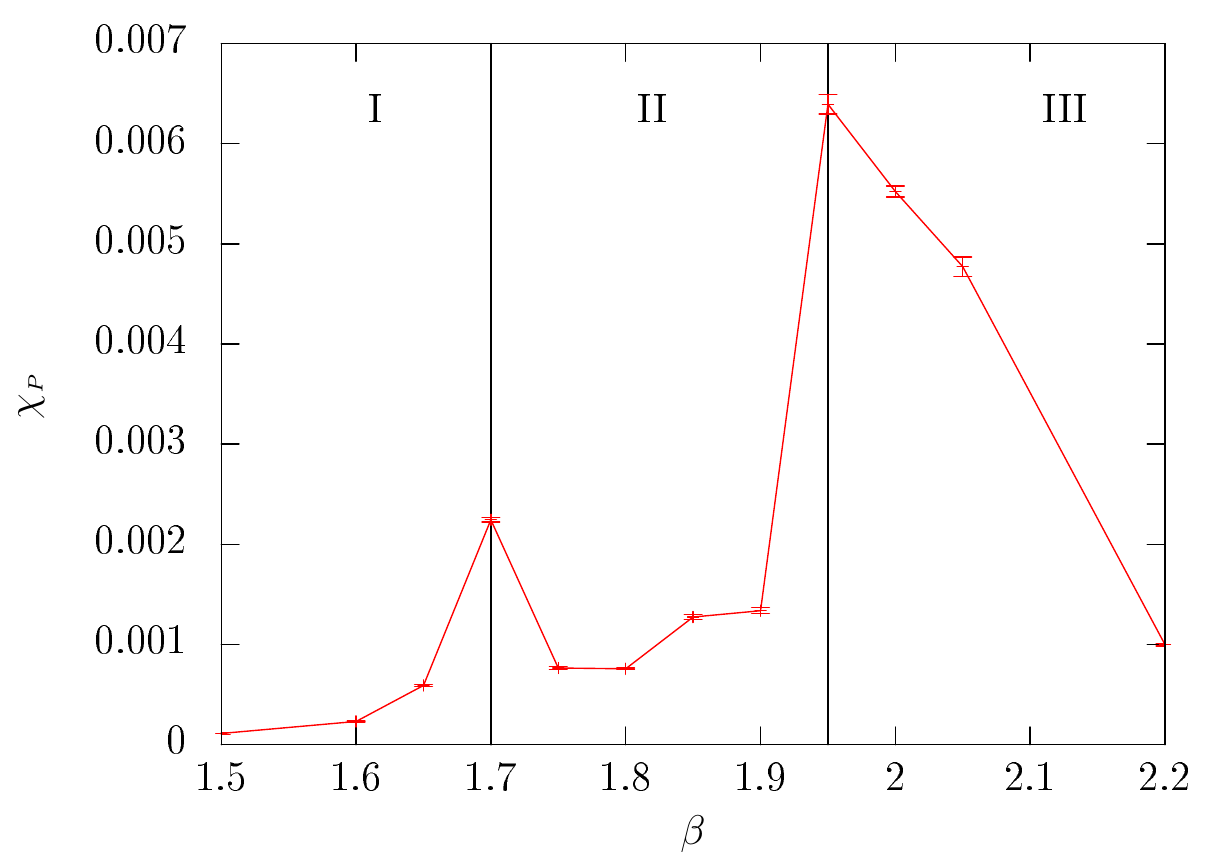}
\caption{The susceptibility of $|P_L|$ is shown as a function of $\beta$ for $\kappa=0.16$ from simulations on a $4\times 16^3$ lattice.
This volume is sufficiently large to identify the transitions as peaks of the susceptibility.
The two distinct peaks indicate two transitions separating three different phases. \Rnum{2} 
has larger values of $\langle|P_L|\rangle$ associated with broken $\Z_2$ centre symmetry and a
deconfined phase. The two other phases are in a confined phase with unbroken centre symmetry.}
\label{fig:bscansus}
\end{figure}
\begin{figure}[t]
\centering
\includegraphics[width=0.7\textwidth]{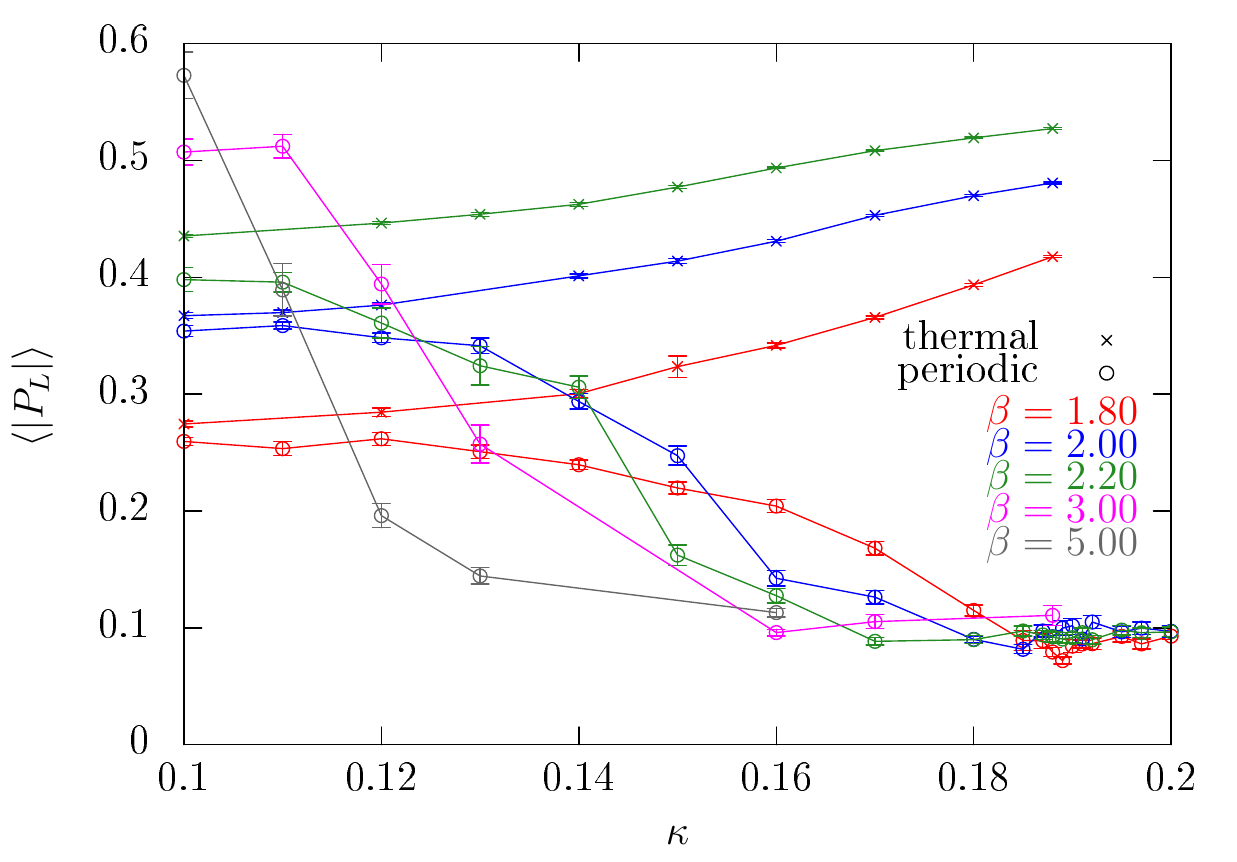}
\caption{Scans of a region of the bare mass parameter $\kappa$ for several fixed values of $\beta$.
In each scan $|P_L|$ is measured in simulations on $4\times 8^3$ lattices. 
Thermal (antiperiodic) and periodic fermion boundary conditions are compared.}
\label{fig:kscan}
\end{figure}
\begin{figure}[t]
\centering
\includegraphics[width=0.7\textwidth]{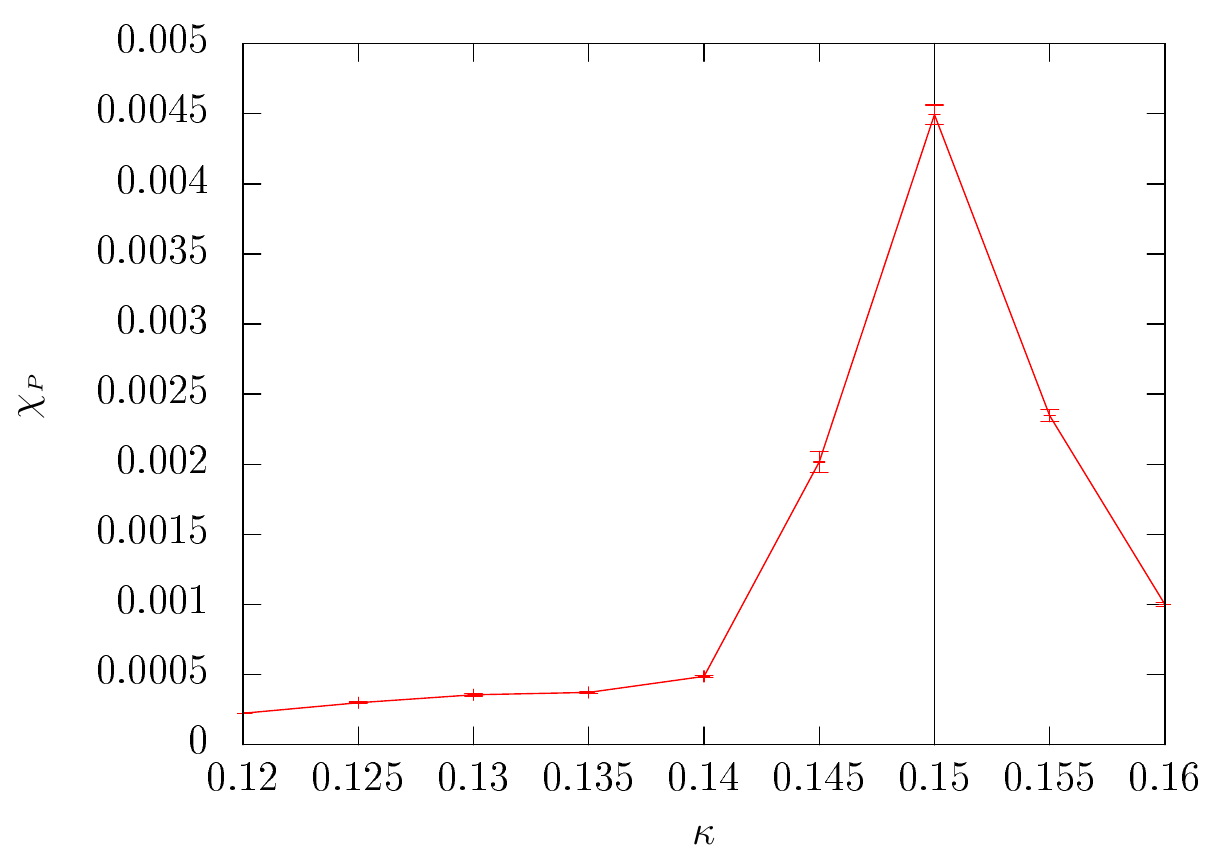}
\caption{The susceptibility of $|P_L|$ at $\beta=2.2$ on a $4\times 16^3$ lattice in a scan of the bare mass 
parameter $\kappa$ with periodic fermion boundary conditions. 
The peak corresponds to the critical $\kappa$ of the transition.}
\label{fig:kscan2}
\end{figure}
\begin{figure}[t]
\centering
\subfigure[\label{fig:histksmalla}]{\includegraphics[width=0.45\textwidth]{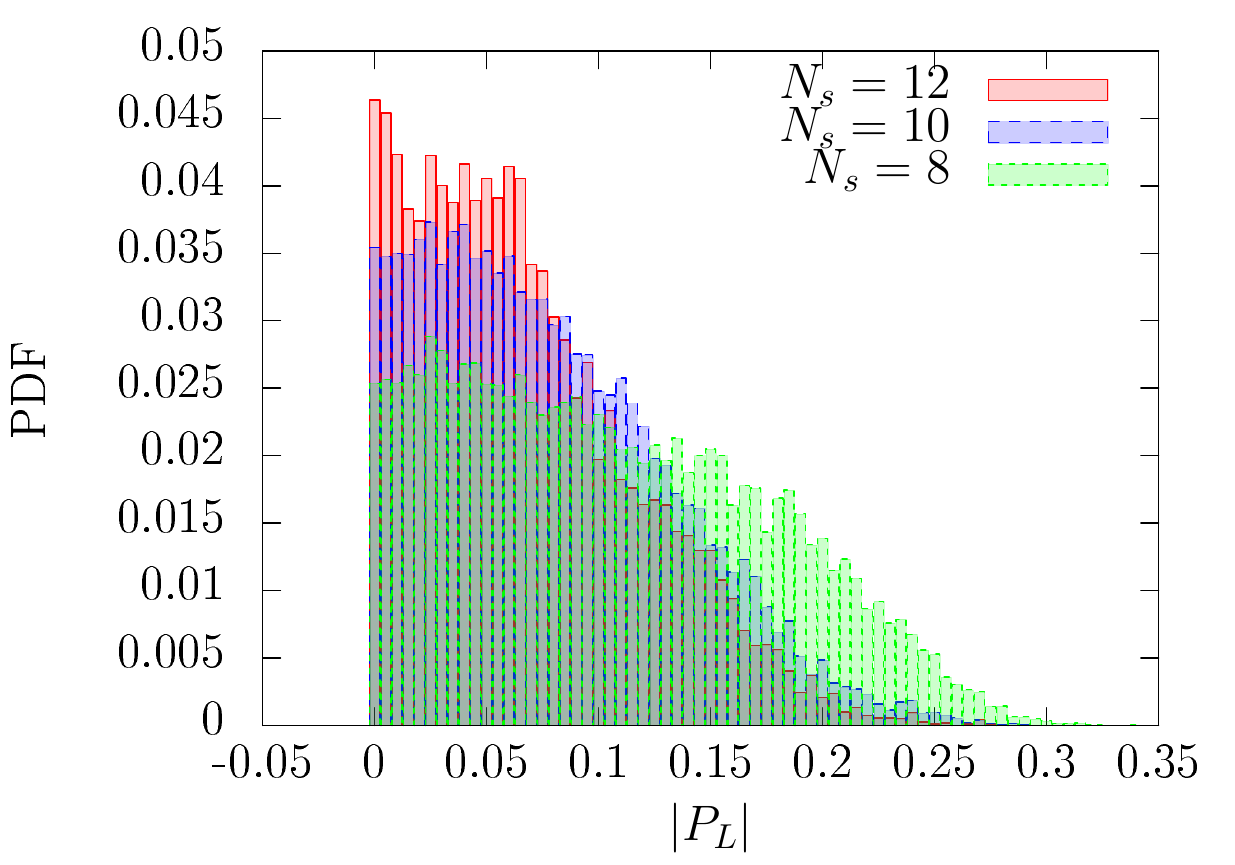}}
\subfigure[\label{fig:histksmallb}]{\includegraphics[width=0.45\textwidth]{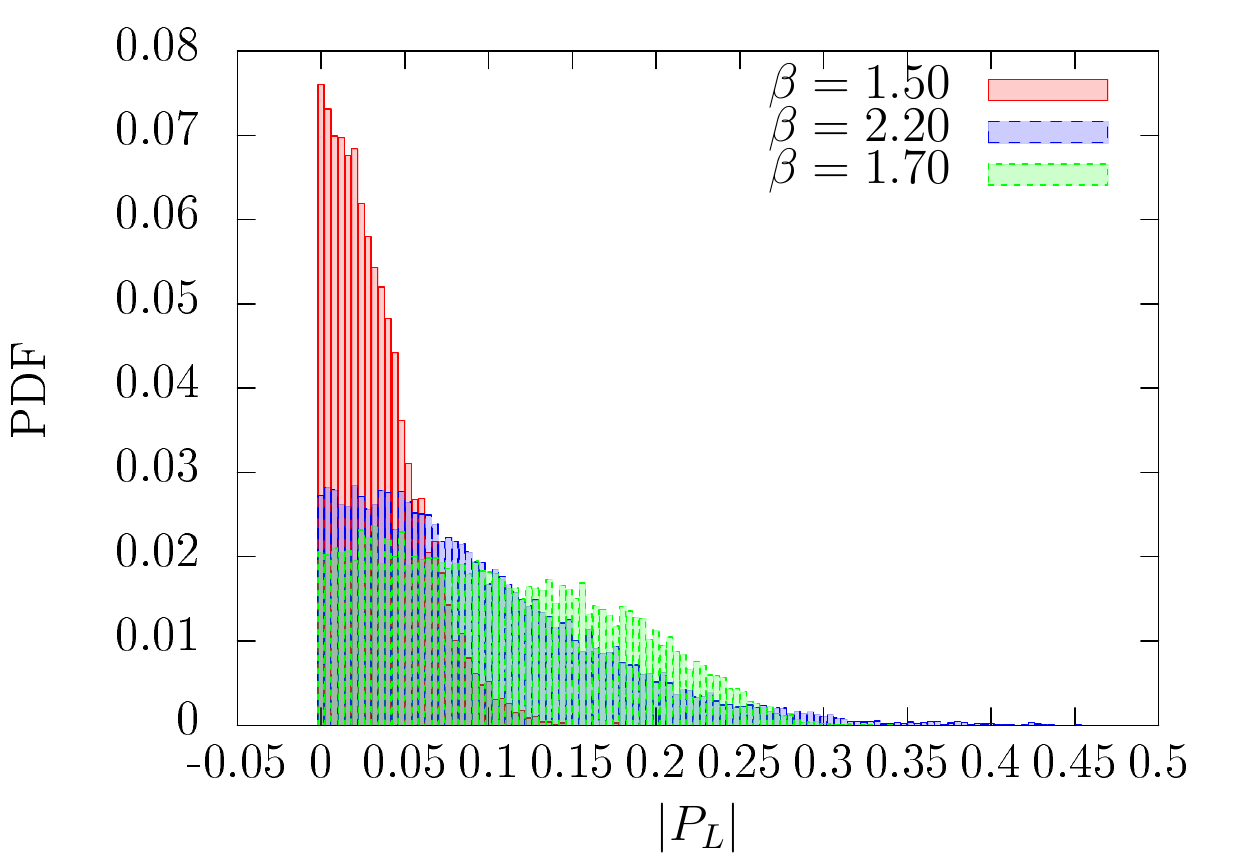}}
\caption{(a) The histogram of $|P_L|$ at $\beta=1.70$ and $\kappa=0.188$.
The chosen $\beta$ corresponds to the maximum value of $\langle|P_L|\rangle$ in Fig.~\ref{fig:bscan} at this bare gluino mass.
Different volumes are compared in simulations on $4\times N_s^3$ lattices, where $N_s=8$, $10$, and $12$.
(b) A comparison of the histograms at $\kappa=0.188$ and different values of $\beta$ and $N_s=8$.
}
\label{fig:histksmall}
\end{figure}
\subsection{The three different phases at a large values of the gluino mass}
We begin our investigations with a scan of the relevant range of $\beta$ values at fixed $\kappa$
and compare the behaviour of the order parameter $P_L$ for the two boundary conditions.
As expected at low $\beta$, corresponding to a large $R$ or low $T$, the theory is confined regardless of 
the boundary condition and the fermion mass. Consistent with our previous investigations we find that 
a decreasing $\bcd$ for smaller bare mass parameters, i.~e.\ larger $\kappa$, in TSYM.

In PSYM the opposite behaviour is observed: the onset of the order parameter shifts towards 
larger $\beta$ values as the bare mass is decreased (see Fig.~\ref{fig:bscan}). We observe that $\langle|P_L|\rangle$
 reaches a maximum at intermediate $\beta$ until it decreases again at large $\beta$. This is a first indication of three different phases: a confined phase at large $R$ connected to the low temperature phase of the thermal theory,
an intermediate deconfined phase, and a second confined, or re-confined, phase at small $R$.

In the re-confined phase $|P_L|$ has a larger expectation value compared to the low temperature confined phase. This is, however, 
not a signal for a deconfined phase. In a deconfined phase the larger expectation value of the modulus of the Polyakov loop indicates a peak of the distribution of the order parameter at $P_L\neq0$, that corresponds to a minimum of the constraint effective potential at this point.

On the other hand, a small rise of the $\langle|P_L|\rangle$ can also be an effect of the modulus function induced by a broadening of the distribution of the order parameter, even though the histogram is peaked at zero. The minimum of the constraint effective potential remains in this case at $P_L=0$, but its curvature at the minimum gets smaller.
We expect a broadening of the distributions at large $\beta$ due to the flat perturbative effective potential.
Different phases can hence be clearly pointed out only from a detailed investigation of the shape of the histogram of 
the order parameter.
 
The best way to distinguish a phase transition from a broadening of the distribution is the 
investigation of finite size effects. Comparing different volumes, we are able to distinguish the broad distribution generated by 
tunnelling between the two $\Z_2$ symmetric minima of the constraint effective potential in the deconfined phase 
and a broad distribution peaked at zero in a confined phase. If the contributions close to zero are suppressed in the histograms
at larger volumes, the theory is in the deconfined phase.

The comparison of the histograms for $\kappa=0.16$ and $\beta=1.8$, $2.0$, and $2.2$ is shown in  Fig.\ \ref{fig:hist}.
These data show a deconfined phase at $\beta=1.8$, a transition close to $\beta=2.0$, and the second confined phase at $\beta=2.2$.
The suppressed tunnelling between the two $\Z_2$ symmetric vacua  for larger volumes at $\beta=1.8$ is clearly visible.
The second confined (re-confined) phase at $\beta=2.2$ is indicated by the distribution around a peaked maximum value at the origin.
At larger volumes the tendency towards one clear maximum at zero is even increased.
Compared to the distribution in the confined phase, the fluctuations in this second confined phase are large, leading to a rather broad distribution. The larger values of $\langle|P_L|\rangle$ in the confined phase at large $\beta$ are hence not a signal for deconfinement; instead they are only indicating this broad distribution.

Different phases can be clearly separated by the peaks in the susceptibility of $P_L$ as shown in Fig.~\ref{fig:bscansus}.
We have found that the separation is only possible at rather large volumes.
The first peak indicates the transition from the confined phase at small $\beta$ to the deconfined phase in correspondence to
the thermal deconfinement transition.
At large $\beta$ there is a second peak separating the deconfined phase from third phase with unbroken centre symmetry.
The transition is characterised by large values for the susceptibility at the peak and in the confined region after the peak.
The large susceptibility reflects the mentioned broad distribution of the order parameter.

\subsection{The transition at a small compactification radius}
We now turn to the transition line at small $R$, i.~e.\ large $\beta$.
The best way to illustrate the transitions in this region are scans of a range of $\kappa$ values at fixed
$\beta$, see Fig.~\ref{fig:kscan}. The chosen values are all above the $\bcd$ of $\SU(2)$ YM.
The dependence of $\langle|P_L|\rangle$ on $\kappa$ illustrates the drastic difference between TSYM and PSYM. 
While for thermal boundary conditions the expectation
value of the order parameter increases as the gluino mass gets smaller, a significant decrease 
is observed in PSYM. This is the signal of the second confined 
(re-confined) phase at larger $\beta$ values.

At very heavy gluino masses the expectation value of the Polyakov loop tends to its pure YM limit and there is 
always a deconfined region close to the $\kappa=0$ axis. 
The boundary  of that region can be identified by the steepest decrease of $\langle|P|\rangle$ as a function 
of $\kappa$ for each $\beta$ and also from the susceptibility (Fig.~\ref{fig:kscan2}).
 Our results depicted in Fig.~\ref{fig:kscan} show that the deconfined 
region shrinks and the transition gets sharper at larger $\beta$ values. Therefore, we conjecture that the transition 
moves from first order at very large $\beta$ towards a crossover at $\beta$ that are smaller, but still above 
the phase transition of YM.
\subsection{Indications for a connection between the two confined phases}
Our results for PSYM show only a mild change of $\langle|P_L|\rangle$ between small and large values of $\beta$ at the smallest gluino mass in Fig.~\ref{fig:bscan} ($\kappa=0.188$).
This change could indicate a transition to an intermediate deconfined phase, but it could also
be due to a mere broadening of the distribution of the order parameter.
A closer investigation of the histograms points towards the latter situation.

The histograms of the data from simulations on a $4\times8^3$ lattice at $\kappa=0.188$ and different $\beta$ (see Fig.~\ref{fig:histksmallb})
never show a two peak structure. We take the point with the largest $\langle|P_L|\rangle$ as a reference for the finite volume analysis. The histogram of the order parameter for three different volumes is shown in Fig.~\ref{fig:histksmalla}. For larger volumes the distribution tends to sharpen around the peak at zero.
There is hence no indication in our results for a transition to a deconfined phase already at $\kappa=0.188$.

This also means that there is a connection between the low $\beta$ and large $\beta$ confined phases. 
Confined and ``re-confined'' phase are in fact one large confined region in phase diagram. 
\section{Conclusions}
\label{sec:conclusions}
We have shown the results of the first lattice simulations of compactified \SU(2) SYM on \RS\ with a soft supersymmetry 
breaking gluino mass term and periodic fermion boundary conditions. In accordance with theoretical predictions, our results clearly point towards the absence of the 
deconfinement transition in the supersymmetric limit. Already at rather large gluino masses we have found no indications 
for the transition in the histograms of the order parameter up to a very small compactification radius.

The deconfinement transition line is even more strongly influenced by the different fermion boundary conditions than suggested by the theoretical predictions \cite{Unsal:2008ch}, that assume a continuity (i.~e.\ absence of deconfinement) only in the supersymmetric limit at zero fermion mass. In addition, an intermediate deconfined phase between two confined regions in the scans at a 
larger bare mass is not predicted for PSYM.  These observations are more consistent with the theoretical predictions for theories with 
a larger number of Majorana fermions than with those for SYM.

Especially the results obtained with a fixed bare coupling constant (Fig.~\ref{fig:kscan}) clearly confirm the difference between the
periodic and antiperiodic fermion boundary conditions and also indicate the connection to the the pure Yang-Mills limit, 
i.~e.\ infinite gluino mass. Close to this limit, there is always a deconfined region for $\beta$ larger than $\bcd$ of YM with a transition to
the confined phase at a certain critical gluino mass. The deconfined region shrinks as $\beta$ is increased. This fact also supports rather the scenario
depicted in Fig.~\ref{fig:theoryphase2} than the one in Fig.~\ref{fig:theoryphase1} for the phase transitions in SYM on \RS.

It is important to note that the finite lattice spacing leads to a breaking of supersymmetry, that invalidates the balance between fermionic and bosonic contributions. 
The breaking is induced by the Wilson mass in the Dirac operator and the violation of the Leibniz rule on the lattice
\cite{Bergner:2009vg}. The rather flat effective potential might be sensitive even to small perturbations by lattice artefacts.
This might explain the observed difference between the measured and predicted transition lines.
Therefore, an important next step is a detailed comparison of different $\Nt$, that corresponds to a study of the theory with finer lattices. 
Nevertheless one expects that the lattice artefacts might have a small impact on the general picture,
in particular on the results at large $\beta$ values.

Besides the most important investigation of the dependence on the lattice spacing, 
further investigations are still required to confirm these results and there are several aspects that we plan to consider in further, more demanding, numerical simulations. 
The scale should be set by measurements of the mass ratios to change the axes of phase diagram from bare parameters to renormalised quantities. The influence of the boundary condition on the
chiral transition line should also be investigated. On large volumes the clear separation of the phases allows in
principle an extrapolation of the transition lines. In this way the critical bare mass for the disappearance of the deconfinement transition can be estimated with a much better precision 
than in our current measurements.

A first exploratory study of our collaboration \cite{Hendrich} considers also the compactification 
of more than one space-time dimension that can relate the results to the investigation of 
finite size effects \cite{Bergner:2012rv}.
\section*{Acknowledgements}
We gratefully thank G.~M\"unster and P.~Giudice for many useful instructions and comments 
and M.~\"Unsal for helpful discussions. The authors gratefully acknowledge the computing 
time granted by the John von Neumann Institute for Computing (NIC) provided on the 
supercomputer JUROPA at J\"ulich Supercomputing Centre (JSC). Further computing time
has been provided by the compute cluster PALMA of the University of
M\"unster and the LOEWE-CSC of the University of Frankfurt. 
 G.B.\ has been supported by the German BMBF, No. 06FY7100.

\end{document}